\documentclass[prd,preprint,eqsecnum,nofootinbib,amsmath,amssymb,
               tightenlines,dvips]{revtex4}

\include{epsf}
\usepackage{graphicx}% Include figure files
\usepackage{bm}% bold math

\def\LL{\Lambda_{\rm Landau}}

\def\q{{\bm q}}

\def\Im{{\rm Im}}

\def\half{{\textstyle\frac{1}{2}}}

\def\cf{C_{\rm F}}
\def\da{d_{\rm A}}
\def\df{d_{\rm F}}

\def\nf{N_{\rm f}}

\def\nc{N_{\rm c}}
\def\PiL{\Pi_{\rm L}}
\def\PiT{\Pi_{\rm T}}
\def\Pivac{\Pi_{\rm vac}}
\def\mD{{m_{\rm D}}}
\def\g2eff{g^2_{\rm eff}}
\def\Tr{\rm Tr \,}
\def\mubar{\overline{\mu}}
\def\muDR{\overline{\mu}_{\rm DR}}
\def\MS{$\overline{\rm MS}$}
\def\nb{n_{\rm b}}
\def\alphas{\alpha_{\rm s}}
\def\sumint{\hbox{$\sum$}^{\rm B}  \!\!\!\!\!\!\!\!\!\!\int}
\def\sumintp{\hbox{$\sum$}^{\rm F} \!\!\!\!\!\!\!\!\!\!\int}

\newcommand\Eq[1]{Eq.~(\ref{#1})}

\def\slashchar#1{\setbox0=\hbox{$#1$}           % set a box for #1 
   \dimen0=\wd0                                 % and get its size
   \setbox1=\hbox{/} \dimen1=\wd1               % get size of /
   \ifdim\dimen0>\dimen1                        % #1 is bigger
      \rlap{\hbox to \dimen0{\hfil/\hfil}}      % so center / in box
      #1                                        % and print #1
   \else                                        % / is bigger
      \rlap{\hbox to \dimen1{\hfil$#1$\hfil}}   % so center #1
      /                                         % and print /
   \fi}                                         %

\advance\parskip 2pt

\begin{document}

%\preprint{APS/123-QED}

\title{Pressure of Hot QCD at Large $\nf$}

\author{Guy D. Moore}
\affiliation
    {%
    Department of Physics, 
    McGill University,
    3600 rue University,
    Montr\'{e}al, QC H3A 2T8,
    Canada
\vspace{1.0in}
    }%
%
%\author{Laurence G. Yaffe}
%\affiliation
%    {%
%    Department of Physics,
%    University of Washington,
%    Seattle, Washington 98195
%    USA
%    }%
%

%\date{\today}

\begin{abstract}
\vspace{0.2in}
We compute the pressure and entropy of hot QCD in the limit of large number
of fermions, $\nf \gg \nc \sim 1$, to next to leading order in
$\nf$.  At this order the calculation can be done exactly, up to
ambiguities due to the presence of a Landau pole in the theory; the
ambiguities are $O(T^8/\LL^4)$ and remain negligible long after the
perturbative series (in $g^2 \nf$) has broken down.  Our
results can be used to test several proposed resummation schemes for the
pressure of full QCD.
\end{abstract}

%\pacs{Valid PACS appear here}


\maketitle

%%%%%%%%%%%%%%%%%%%%%%%%%%%%%%%%%%%%%%%%%%%%%%%%%%%%%%%%%%%%%%%%%%%%%%%%%%%%%%%

\section {Introduction}

At temperatures large compared to the intrinsic scale of QCD,
$T \gg \Lambda_{\rm QCD}$, the running QCD coupling constant
is small, and perturbation theory should be a useful tool for studying
thermodynamic and dynamic properties of the theory.  This is true in
principle, at least.  

In practice, there are obstacles to the use of perturbation theory.
The first is that hot QCD has an intrinsically
nonperturbative scale, at wave numbers $k \sim g^2 T$, or length scales
$l \sim 1/g^2 T$.  (The running coupling $g^2$ should be evaluated at a
scale of order $T$.)  The presence of nonperturbative physics at this
scale has nothing to do with asymptotic freedom or the existence of the
scale $\Lambda_{\rm QCD}$; even at temperatures many orders of magnitude
above $\Lambda_{\rm QCD}$, nonperturbative physics sets in at a scale
only a little lower than the thermal scale, 
$g^2 T \sim T / \log(T/\Lambda_{\rm QCD})$.  Some observables
are not very sensitive to this infrared scale; but any observable will
be sensitive to this scale at some order in the coupling expansion.
For instance, the Debye screening length carries nonperturbative
corrections which are suppressed by only one power of $g$
\cite{Rebhan_mD,ArnoldYaffe_mD}; photon emissivity is believed to receive
nonperturbative corrections suppressed by $g^2$ \cite{powercount}; and
the baryon number violation rate arises entirely from nonperturbative
physics,and is sensitive at leading order \cite{ASY_etc}.

The measurables which show thermal effects with the least sensitivity to
the nonperturbative scale are the standard thermodynamic parameters,
such as pressure, entropy density, and energy density.
For these quantities, perturbation theory can
determine all coefficients up to $O(g^5)$, and that calculation has been
carried out \cite{ArnoldZhai,ZhaiKastening,BraatenNieto}.
(It is also possible to compute the $O(g^6 \ln(g))$ term, and this is in
progress \cite{York}.)  With the coefficients of the pressure up to
$O(g^5)$ in hand, one can diagnose the quality of the perturbative
expansion, by seeing how quickly successive partial sums converge.  Here
we meet the other problem of perturbation theory at finite temperature;
even when the perturbative series exists to some order, it shows very
poor convergence.  This problem is distinct from the presence of
nonperturbative magnetic physics; for instance it also appears in hot QED,
which is free of strongly coupled IR behavior \cite{QED,ArnoldZhai}.

Recently this problem has received renewed attention.  Three groups have
proposed reorganizations or partial resummations of the thermal
perturbative expansion, which are hoped to improve the convergence, and
therefore the utility, of perturbation theory.  Andersen,
Braaten, and Strickland, following an earlier proposal by Karsch {\it et
al.\ } in scalar field theories \cite{Karschetal},
have proposed to do so by adding the hard
thermal loop (HTL) Lagrangian to the action at tree level and
subtracting it again with a counterterm treated as formally higher
order, a technique they call hard thermal loop perturbation theory
\cite{ABS1,ABS2,ABS3}.  Blaizot, Iancu, and Rebhan have proposed a
technique based on 2PI $\Phi$ derivable methods, together with a set of 
approximations to render the method tractable, also based on hard
thermal loops \cite{BIR1,BIR2,BIR3}.  A similar proposal was recently
made by Peshier \cite{Peshier}.

All three groups have produced predictions for the QCD pressure, both
for pure glue and for QCD with fermions.  Unfortunately, it is difficult
to judge whether the techniques are successful, since all we know
independently about the QCD pressure are the perturbative series, which
is only useful at uninterestingly weak coupling, and lattice results
from the Bielefeld group \cite{Karsch}, which are only at such large
coupling that it is not clear whether the resummation techniques should
still work.  The resummation techniques have also been applied to scalar
field theory, but in that case exact results are only known for O($N$)
scalar theory at large $N$, which is an almost trivial theory.

For this reason we think it advantageous to have accurate results for
the pressure, in a theory as close to QCD as possible.  In this paper we
present a calculation of the thermal pressure in QCD with a large number
of fermions, $\nf \gg \nc \sim 1$, at next to leading order (NLO) in a
large $\nf$ expansion and for general $g^2 \nf$, which is treated as
$O(1)$. This theory contains a great deal of the 
physics of QCD; for instance, the gauge fields are screened by the
plasma with the same complicated frequency and momentum dependence as in
full QCD.  The perturbative series for the pressure shows poor behavior,
as in full QCD.  Of course, some of the interesting physics of full QCD
is absent at NLO in this expansion.  This is unfortunate;
but on the other hand it is what allows us to solve this theory.
For previous exact results in 6 dimensional scalar theory, see
\cite{BLNR} (who encountered some similar issues, regarding Landau
singularities, to those present here).

We should emphasize now that, since all of the physics of large $\nf$
QCD is also present in full QCD, any approximation which shows poor
performance in large $\nf$ QCD cannot be expected to be valid for full
QCD either.  The reverse need not be the case; an approximation which
works at large $\nf$ may stumble over the physics which is left out in
the large $\nf$ expansion.  Therefore, an approximation's performance
against our results should be considered an optimistic estimate of its
performance in full QCD.

In the next section, we will explain what we mean by large $\nf$ QCD.
In Section \ref{sec:compute}, we describe how the calculation of the QCD
pressure is carried out in this theory.  Our results and conclusions are
presented in Section \ref{sec:results}, but we very briefly summarize
them here.  At small effective coupling $g^2 \nf$, the pressure is close
to the free theory value, and perturbation theory works well.
As the effective coupling is increased, $P/T^4$ falls at first, but
eventually rises; it becomes of order its free theory value about where
Landau pole ambiguities become uncomfortably large.  The entropy shows
the same behavior.  It is difficult to obtain such behavior in an ideal
quasiparticle approximation, such as the model of Peshier, K\"{a}mpfer,
Pavlenko, and Soff \cite{Peshier_old}.

\section{Large $\nf$ QCD}
\label{sec:Nf}

By large $\nf$ QCD, we mean QCD with the number of colors $\nc$ taken as
fixed and $O(1)$, but with the number of fermionic species $\nf$ taken
large and the gauge coupling $g^2$ taken small, so that the combination
$g^2 \nf$, which serves as an effective coupling strength, remains fixed
and $O(1)$.  Therefore, in $\nf$ counting, a factor of $g^2$ is treated
as $O(1/\nf)$.  This procedure is appropriate because it is the
effective coupling which determines the convergence of perturbation
theory.

In the following it will be useful to define $\g2eff$ as
\begin{equation}
\g2eff \equiv \frac{g^2 \nf \cf \df}{\da} = \left\{
	\begin{array}{cc} 
	\displaystyle \frac{g^2 \nf}{2} \, , & {\rm QCD} \, , \\ & \\
	g^2 \nf \, , & {\rm QED} \, . \\ \end{array} \right.
\end{equation}
Here $\cf$ is the quadratic Casimir of the representation containing the
fermions and $\df$ and $\da$ are the dimensions of the fermionic and
adjoint representations.  In SU($\nc$) gauge theory but with 
adjoint fermions, $\g2eff = g^2 \nf \nc$.
Other than overall factors involving $\nc$, all the nontrivial coupling
dependence of the pressure and entropy which we will find will depend on
$\g2eff$.  Note that $\g2eff$ is renormalization point dependent; at
leading order in $\nf$ its renormalization point dependence is,
\begin{equation}
\label{eq:beta}
\beta(\g2eff) \equiv \frac{\mu \, d \g2eff}{d\mu} = 
	\frac{(\g2eff)^2}{6 \pi^2} \, .
\end{equation}
There are no corrections involving higher powers of $\g2eff$; all
corrections to this relation are suppressed by at least one power of
$1/\nf$.  Therefore, one can solve completely for the scale dependence
of $\g2eff$, at leading order in $1/\nf$;
\begin{equation}
\frac{1}{\g2eff(\mu)} = \frac{1}{\g2eff(\mu')} 
	+ \frac{\ln(\mu'/\mu)}{6 \pi^2} \, .
\label{eq:scale_dep}
\end{equation}
Clearly the theory has a Landau pole at $\LL \sim \mu e^{6 \pi^2/\g2eff}$.
We will define $\LL$ so that the gauge field propagator diverges
at $Q^2 = \LL^2$; under this definition, $\LL=\mu e^{5/6} e^{6 \pi^2/\g2eff}$.
There is no hope that some strong coupling dynamics somehow generate a
strongly coupled UV fixed point, because \Eq{eq:scale_dep} is exact at
leading order in $1/\nf$.  Therefore the theory technically does not
exist. 

This problem is fatal if $\LL \sim T$; but for $\LL \gg T$ we argue that
it is not.  The Landau pole means that the theory as such becomes ill
defined; but it is possible to ``rescue'' the theory by introducing
either additional heavy degrees of freedom or high dimension operators,
if they arise at a scale $\leq \LL$.
For instance, if the operator $F_{\mu \nu} D^2 D^2 F^{\mu \nu}/\LL^4$ is
added to the Lagrangian with coefficient greater than 1, the gauge
propagator becomes nonsingular at all energies, which is sufficient for
our NLO calculation to proceed.  
% For those of you reading this at home, wondering why I use two powers
% of $D^2$; one power can render the propagator well behaved EITHER for
% timelike or spacelike 4-momenta but not both.
This term ruins the
renormalizability of the theory, but in a way which will not appear
in the NLO calculation of the pressure.  What the Landau pole does mean is,
that the exact definition of the theory is ambiguous, and quantities
like the entropy or the pressure are only completely well defined after
the physics which resolves the Landau singularity has been specified.
The size of the ambiguity is, however, suppressed by a positive power of 
$(T/\LL)$, which for the pressure at NLO turns out to
be $(T/\LL)^4$.  So provided that $\LL$ is kept suitably larger than
$T$, the relative size of the ambiguity is tiny and can be ignored.%
\footnote
    {%
    Similar considerations, in the context of scalar field theory,
    appear in Drummond {\it et al.}, \cite{DHLR}.
    }
As $\g2eff$ is increased, this new ambiguity remains tiny long after the
convergence of perturbation theory for the pressure has broken down.
Therefore, we will forge ahead and compute the pressure, without
specifying the UV completion; but we will only work with values of
$\g2eff(\mu)$ for 
which $(\LL/T) > 35$, which proves abundantly sufficient to keep the
relative ambiguity in the NLO pressure below $1\%$.

The other feature of large $\nf$ gauge theory which we should emphasize
is, that the nonabelian nature of the theory is not very important.
This is because the gauge field self-interactions carry powers of $g$
without introducing powers of $\nf$.  In particular the gauge
self-interactions will have no relevance in the calculation we perform
here, and first appear, in the pressure and entropy, at NNLO (next to
next to leading order) in $1/\nf$.  Therefore our results will apply
both to QED and QCD.

\section{Computation of the Pressure}
\label{sec:compute}

In a relativistic field theory with vanishing chemical potentials, the
pressure and free energy density are opposites, $P=-F$; so we compute
the pressure by computing the free energy, which equals the trace of the
sum of all 1PI vacuum bubble graphs.  This is to be done at finite
temperature, which means in Euclidean space with time extent
$\beta=1/T$, periodic boundary conditions for bosons, and antiperiodic
boundary conditions for fermions.  To get the {\em thermal} pressure, we
subtract off the result at zero temperature; the difference will be
finite, though renormalization is still necessary because the result
depends on the coupling $\g2eff$ and this coupling runs.  

\begin{figure}[t]
\centerline{\epsfxsize=5.5in\epsfbox{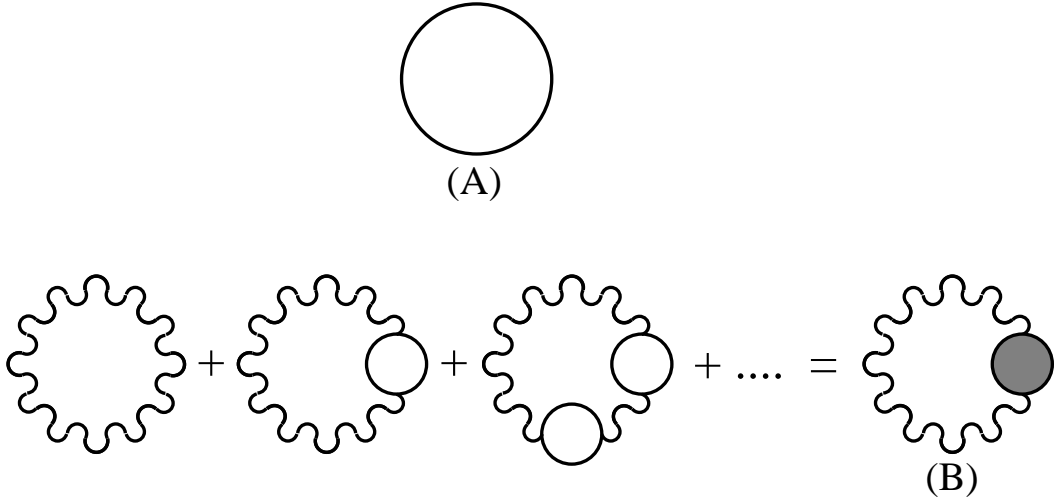}}
\caption{Leading order, $O(\nf)$ diagram, (A), which is a free fermion
bubble, and NLO, $O(\nf^0)$ diagrams, (B), which are a gauge boson loop
with any number of fermion bubble self-energy insertions.  (Coupling
counterterm insertions, though not shown, should also be resummed.)
They can be resummed by the usual Schwinger-Dyson method, to give a loop
with a resummed gauge field propagator. \label{fig1}}
\end{figure}

At leading order, $O(\nf)$, there is only one diagram, a bare fermionic
loop, shown in Fig.~\ref{fig1} (A).  Its contribution to the pressure
is%
\footnote
    {%
    We use a [++++] Euclidean and [ - +++] Minkowski metric.
    Four-vectors are written with upper case letters, their
    space and time components in lower case, so if $Q$ is a 4-vector,
    then $q^0$ is its time component, $\q$ is the
    spatial vector and $q\equiv |\q|$ its magnitude.  The Minkowski
    continued frequency is typically denoted with a different letter,
    usually $\omega$.
    }
\begin{equation}
P_{\rm LO} = \nf \, \Tr \left( \sumintp_{Q} \; \, \ln \slashchar{Q} \; 
	- \int_{Q} \, \ln \slashchar{Q} \right)
\end{equation}
Here the trace is over group and Dirac indices, and
the fermionic sum-integral symbol means
\begin{equation}
\sumintp_{Q} \; \equiv T \!\!\!\!\!\!
	\sum_{q^0 = (2n+1)\pi T}^{n \in {\cal Z}}
	\int \frac{d^3 \q}{(2\pi)^3} \, ; \qquad
\int_{Q} \equiv \int \frac{d^4 Q}{(2\pi)^4} \, .
\end{equation}
We have written this in 3+1 dimensions, but dimensional regularization
is implied.  At leading order, since the $T=0$ value is subtracted off,
this is irrelevant, the result is finite and equals
\begin{equation}
P_{\rm LO} = 4 \nf\, \df \: \frac{7}{8} \> \frac{\pi^2 T^4}{90} \, .
\label{eq:LO}
\end{equation}
This is the usual free theory value; $4 \nf \df$ counts the number of
degrees of freedom [the 4 counts the two spin states each for particle
and anti-particle; $\df$ 
is the number of colors, $\df=\nc$ for fundamental representation
fermions in SU($\nc$)].

At NLO, there are an infinite number of diagrams, but the structure is
extremely simple; all contributions look like a gauge boson propagator
with an arbitrary number of fermion loops, and counterterm insertions, 
inserted.  No other graphs
contribute at $O(\nf^0)$.  This is because the only way to add a loop to
a graph without changing the power of $\nf$, is to add a fermion
self-energy loop (or the required counterterm) to a gauge boson propagator;
any other way to add a loop introduces a $g^2$ without another factor of
$\nf$, and $g^2 \sim 1/\nf$.  This is the reason that large $\nf$ gauge
theory is so simple; the large $\nf$ limit is a much stronger
organizing principle than large $\nc$.

This set of diagrams can of course be resummed by the standard
Schwinger-Dyson trick.  Therefore the NLO contribution to the pressure
is given by
\begin{equation}
P_{\rm NLO} = - \frac{1}{2} \Tr \left( \sumint_{Q}
	\; \ln \left[ [G^{-1}_0]^{\mu \nu}(Q) 
	+ \Pi^{\mu \nu}_{\rm th}(Q) \right] \; \;
	- \int_{Q} \, \ln \left[ [G^{-1}_0]^{\mu \nu}(Q) 
	+ \Pi^{\mu \nu}_{\rm vac}(Q) \right]
	\right) \, ,
\label{eq:PNLO}
\end{equation}
where the trace runs over group and Lorentz indices.  The group trace is
trivial, it gives $\da$ [which is $\nc^2-1$ for SU($\nc$) and 1 for
QED].  The bosonic sum-integral symbol means
\begin{equation}
\sumint_{Q} \; \equiv T \!\!\!\!
	\sum_{Q_0 = 2n\pi T}^{n \in {\cal Z}}
	\int \frac{d^3 q}{(2\pi)^3} \, ,
\end{equation}
$\Pi^{\mu \nu}_{\rm th}$ is the fermionic plus counterterm 
contribution to the gauge
boson self-energy at finite temperature, 
$\Pi^{\mu \nu}_{\rm vac}$ is the vacuum value, and $G^{-1}_0$ is the free
inverse propagator.  Our sign convention for $\Pi$ is chosen so that
our $\Pi$ agrees with the usual Minkowski, [+ - - -] $\Pi$.
The same expression for the pressure can be derived from the
Luttinger-Ward relation \cite{2PI}, with a comparable amount of work.

Both vacuum and thermal  self-energies are gauge invariant, since their
evaluation doesn't involve gauge field propagators.  They are also 
exactly transverse, $Q_\mu \Pi^{\mu \nu}(Q)=0$, and together with
rotational invariance this ensures that 
$\Pi^{\mu \nu}_{\rm th}-\Pivac^{\mu \nu}$ can be
expressed in terms of two scalar functions \cite{Weldon},
\begin{equation}
\Pi_{\rm th}^{\mu \nu}(Q)-\Pivac^{\mu \nu}(Q) 
= P^{\mu \nu} \PiT(Q) + Q^{\mu \nu} \PiL(Q) \, .
\end{equation}
The exact form of $P^{\mu \nu}$ and $Q^{\mu \nu}$ are not important for
our problem, all that matters to us is that they are properly normalized
projection operators, $P^\mu_\mu=2$, $Q^\mu_\mu=1$, and 
$P^{\mu \nu}(Q) + Q^{\mu \nu}(Q) + Q^\mu Q^\nu/Q^2 = \eta^{\mu \nu}$
(which is $\delta^{\mu \nu}$, since right now we are in Euclidean
space).  
For the vacuum self-energy, the transverse and longitudinal components
are the same, and their value is well known;
\begin{equation}
\Pivac^{\mu \nu} = (P^{\mu \nu}+Q^{\mu \nu}) \, \Pivac \, ,
\qquad
\Pivac = - \frac{\g2eff\, Q^2}{12\pi^2} \left(
	\ln \frac{Q^2}{\mubar^2} - \frac 53 \right) \, .
\end{equation}
Here $\mubar$ is the \MS\ renormalization point.

Some gauge fixing prescription is necessary to evaluate \Eq{eq:PNLO},
though the result is gauge invariant since it represents 
a complete treatment to some order in a systematic expansion (this can be
explicitly checked).  
If we choose a gauge where $G^{-1}_0$ factorizes in the same
way as $\Pi$, the Lorentz trace and logarithm will commute.  For instance, in
Feynman gauge, $[G^{-1}_0]^{\mu \nu}(Q) = \eta^{\mu \nu} Q^2$, and
(dropping the overall group factor)
\begin{equation}
\Tr \ln \left[ [G^{-1}_0]^{\mu \nu}(Q) 
	+ \Pi^{\mu \nu}_{\rm th}(Q) \right]
	= 2 \ln (Q^2 + \PiT + \Pivac) + \ln(Q^2 + \PiL + \Pivac) 
	+ \ln(Q^2) - 2 \ln(Q^2)
	\, .
\end{equation}
The contribution containing $\PiT$ comes from $P^{\mu \nu}$; the
contribution containing $\PiL$ comes from $Q^{\mu \nu}$; the positive
$\ln(Q^2)$ contribution comes from $Q^\mu Q^\nu/Q^2$, and the
$-2 \ln(Q^2)$ contribution comes from the ghosts, which have no
self-energy at this order in $\nf$.  In a general covariant gauge the
$Q^\mu Q^\nu/Q^2$ term would be shifted by $\ln(\xi)$ with $\xi$ 
the gauge fixing
parameter; but this contributes equally to the vacuum and thermal
contributions and so drops out.%
\footnote
    {%
    This is because $T \sum_{q^0} C = \int (dq^0/2\pi) C$ for any $C$
    which is $q^0$ independent.  The easiest way to see this is directly
    in the time domain, though it also follows easily on rotating to
    Minkowski frequency.  The result for $P_{\rm NLO}$ in
    Coulomb gauge differs by a factor of $\ln(q^2)$ from the covariant
    result, but this also cancels between vacuum and thermal for the
    same reason.%
    }
The next task is to evaluate the thermal self-energy.  This has been
done by Weldon \cite{Weldon} (in Minkowski space, but the continuation 
is elementary).  His result is listed in the Appendix; unfortunately it
involves an integration which to our knowledge cannot be done in closed
form. 

\begin{figure}[t]
\centerline{\epsfxsize=3in\epsfbox{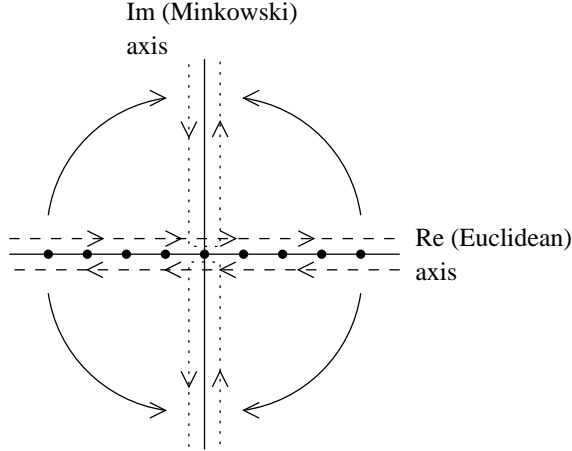}}
\caption{\label{fig:contour} 
Points where summation occurs (dots on Euclidean axis)
become poles whose residues must be summed.  The dashed contour sums
these poles, and can be continued (arrows) to the dotted contour, which
evaluates the discontinuity across the imaginary (Minkowski) axis. }
\end{figure}

Because the self-energies $\PiL$, $\PiT$ are only expressed as single
integrals and not as closed form, analytic expressions, at this point we
must either make approximations, or turn to semi-numerical methods.  We
want exact results; otherwise there is no point in our looking at this
theory.  So the calculation will have to become more
numerical.  The difference between a sum-integral and an
integral is an inconvenient quantity for numerical analysis, especially
when issues of UV regulation have not all been addressed.  At large
values of $Q^2$, the argument in the integral in \Eq{eq:PNLO} is smooth,
and the difference between a sum-integral and an integral is
exponentially suppressed, by $O(\exp(-Q/T))$.  So it is only the moderate
$q^2$ region where we need to handle sum-integration.  In this region,
this can be done by re-expressing the frequency sum as a contour
integration and continuing to Minkowski frequency.  Namely, we note that
the function $\cot(q^0/2T)$ has poles at $q^0 = 2n\pi T$, of residue
$2 T$.  So the Matsubara sum can be replaced by a contour integral, 
\begin{equation}
\sum_{q^0=2\pi nT}^{n \in {\cal Z}} \ln(Q^2 + \Pi(Q))
	= i \int_{C} \frac{d q^0}{4\pi} \cot(q^0/2T) \ln(Q^2 + \Pi) \, ,
\end{equation}
where the contour runs just above the real axis from $-\infty$ to
$+\infty$, and then just below the axis back to $-\infty$.  This is
illustrated in Fig.~\ref{fig:contour}, which also shows that this
contour can be deformed to pinch the imaginary (Minkowski) axis.  The
result from upper and lower half planes are equal, so the contour
integral becomes
\begin{equation}
\sum_{q^0=2\pi nT}^{n \in {\cal Z}} \ln(Q^2 + \Pi(Q))
	= - \int_0^\infty \frac{d\omega}{\pi} 
	{\rm coth}(\omega/2T) \; \Im \: \ln \left( q^2 - \omega^2 
	+ \Pi(q,i\omega+0) \right)
	\, .
\end{equation}
Note that ${\rm coth}(\omega/2T) = 1 + 2/(e^{\omega/T}-1) =
2(\nb(\omega)+1/2)$.  The $T \rightarrow 0$ limit is trivial; just leave
out the Bose distribution and use the vacuum self-energy.
Therefore, \Eq{eq:PNLO} becomes
\begin{eqnarray}
\!\!\!\!\!\!\!\!\!\! P_{\rm NLO} \!\! & = & \!\! \da \!\! \int \! 
	\frac{d^3 q}{(2\pi)^3} 
	\! \int_0^\infty \! \frac{d\omega}{\pi} \bigg[
	2 \left( \vphantom{\frac 12}
	[\nb{+}\half] \Im \, \ln \left(q^2{-}\omega^2{+}\PiT{+}\Pivac\right)
	- \half \Im \, \ln \left(q^2 {-} \omega^2 {+} \Pivac\right) \right)
	\nonumber \\ && \hspace{1.23in}
	+ \left( [\nb{+}\half] \Im \, 
	\ln \frac{q^2{-}\omega^2{+}\PiL{+}\Pivac}{q^2{-}\omega^2}
	- \half \Im \, \ln \frac{q^2 {-} \omega^2 {+} \Pivac}
	{q^2 {-} \omega^2} \right) \! \bigg] \, .
\label{eq:P_minkowski}
\end{eqnarray}
Here $\Pi$ is always understood to be evaluated at $(q,i\omega+0)$.
Note that $\Im \, \Pivac = 
(\omega^2{-}q^2)(\g2eff/12\pi) \theta(\omega^2{-}q^2)$ is
nonzero for timelike momenta; the imaginary parts of the logarithms are
always at least $O(\g2eff)$ away from $\pi$.
There are no infrared problems in
evaluating the above expression; we discuss ultraviolet issues next.

The thermal and vacuum contributions fail to cancel for two reasons.
First, there is the factor of $\nb(\omega)$ associated with the thermal
contribution, which can be traced to the difference between $\sum_{q^0}$
and $\int dq^0$.  There are no ultraviolet problems associated with this
piece, since at large $\omega$ it is exponentially suppressed by
$O(\exp(-\omega/T))$, while at large $q$ but moderate $\omega$, the
thermal self-energy has an exponentially small imaginary part,
$\Im \Pi \sim \exp(-(q{-}\omega)/2T)$.  Therefore this piece is
exponentially ultraviolet safe.

Next, there are $\PiT$ and $\PiL$.  
They are only power suppressed at large
$Q$, and can potentially cause more trouble.  To understand this region
it is best to go back to the Euclidean expression,%
\footnote
    {%
    We should perform pieces linear in $\nb$ in Minkowski space; but
    terms without $\nb$ can be performed either in Minkowski space, or
    in Euclidean space--integrating, not summing, on $q^0$--or using 
    some mixture, as convenient.%
    }
and expand in small $T^2/Q^2$.  
In this small $T^2/Q^2$ limit, the logarithms can be expanded, 
\begin{equation}
2\ln \frac{Q^2 {+} \Pivac {+} \PiT}{Q^2 {+} \Pivac}
+ \ln \frac{Q^2 {+} \Pivac {+} \PiL}{Q^2 {+} \Pivac}
	= \frac{2 \PiT {+} \PiL}{Q^2 {+} \Pivac} 
	- \frac{2(\PiT)^2 {+} \PiL^2}{2 (Q^2 {+}
	\Pivac)^2} + \ldots \, .
\label{eq:expand_log}
\end{equation}
Using \Eq{eq:G} and \Eq{eq:G_expand} from the Appendix, the dominant
large $G$ term is
\begin{equation}
\frac 12 \int \frac{d^4 Q}{(2\pi)^4}
\frac{2\PiT{+}\PiL}{Q^2 {+} \Pivac}
\simeq \frac{7 \pi^2 T^4 \g2eff}{45} \int \frac{d^4 Q}{(2\pi)^4}
\frac{3 (q^0)^2 - q^2}{(Q^2)^3} 
\, \frac{1}{1- \g2eff \ln[(Q^2)/\mubar^2]/12 \pi^2} \, .
\end{equation}
This term is potentially logarithmically divergent.  If one performs an
angular average over the (Euclidean) direction of $Q$, at fixed $|Q^2|$,
then $( 3(q^0)^2-q^2)$ averages to zero, while the rest is a function of
$Q^2$ only; so any regularization or cutoff procedure which respects
Euclidean invariance will be UV well behaved.  However,
if we perform the $q^0$ integration first, at fixed $q$, and then
perform the $d^3 q$ integration, then because of the logarithm in the
denominator, the result does not vanish; instead
the $q^0$ integral gives a result of order $(\g2eff)^2 T^4 / q^3$.
Therefore, performing the $q^0$ (or Minkowski $\omega$) integration
first and the $d^3 q$ integration last holds out the potential for
fake logarithmic divergences.

Any sensible way to UV complete the theory, to deal with the Landau
pole, will respect Euclidean
invariance.  Therefore, however we deal with very large $Q^2$ values, we
should do so in a Euclidean invariant way.  We choose
to apply a large momentum cutoff procedure;%
\footnote
    {%
    The reader may worry that this will introduce gauge fixing
    dependence.  It does not, within the class of Lorentz and Coulomb
    gauges, presumably because the self-energy is gauge invariant;
    recall that the nonabelian nature of the theory is not relevant at
    the order of interest.%
    }
we stop
the $d^4 Q$ integration at $Q^2 = a \LL^2$, for $a<1$, varying the value
of $a$ between 1/4 and 1/2 to estimate the irreducible ambiguity.
Applying a Euclidean invariant high dimension operator cutoff works
similarly. 

As we just indicated, it is necessary to cut off the integration at a
finite value of $Q^2$, lower than $\LL^2$.  This also requires some
extra care with our Minkowski continuation; we must compute the great
arcs at complex $q^0$ which connect the Minkowski and Euclidean
contours.  Our procedure is this.  First, we perform the Minkowski
integration up to $\omega < \LL \sqrt{a}$ and $q < q_{\rm max} \sim 25
T$ (the contribution linear in $\nb$ is carried to larger $q$).  
The Minkowski integration is done by numerical
quadratures, using adaptive mesh refinement.  Great care must be taken
close to the light-cone at large $q$.  To speed things up, the integrals
required for 
the self-energies are expressed in terms of a handful of integrals
of one variable (either $[\omega{-}q]$ or $[\omega{+}q]$), which are
tabulated and spline interpolated so only two integrations nest.
Then we add the great arc from the Minkowski to the Euclidean frequency
axis, and integrate over Euclidean $q^0$ down to 
$(q^0)^2 + q^2 = q^2_{\rm max}$.  This accounts for all $Q^2 < q^2_{\rm
max}$.  Then we integrate over Euclidean $q^2_{\rm max} < Q^2 < a \LL^2$,
performing first the angular integration and then the integration over
the magnitude of $Q^2$.  Note that $(2\PiT{+}\PiL)$ in
\Eq{eq:expand_log} vanishes on angular averaging at least to $O(T^{10})$.
The lowest order term at large $Q^2$, after angular averaging, comes from 
$(\PiT^2 {+} \PiL^2)^2/(Q^2{+}\Pivac)^2 \sim T^8$, so the irreducible
Landau pole introduced ambiguity is $O(T^8 / \LL^4)$, as claimed earlier.

An alternative procedure is to separate the $\nb$ and $1/2$ pieces of
\Eq{eq:P_minkowski} and rotate the $1/2$ piece back to Euclidean space.
The $\nb$ piece is then performed in Minkowski space, where it is
exponentially UV safe; the $1/2$ piece gives \Eq{eq:PNLO} but with the
frequency summation replaced by a frequency integration; this can
directly be done in Euclidean space, rather than by the rather elaborate
contour just described.%
\footnote%
    {%
    This simpler procedure was pointed out to me by Ipp and Rebhan.
    }
The answers agree.

We compute the pressure at a series of values of $\g2eff$, always at
renormalization point
\begin{equation}
\muDR \equiv \pi e^{-\gamma_E} T \simeq 1.76 T\, ,
\end{equation}
with $\gamma_E = 0.577215\ldots$ Euler's constant.  This is the value
suggested by dimensional reduction \cite{KLRS}; at this renormalization
point, the energy density in an infrared magnetic field equals its tree
level value, $\int d^3 x \; B^2/2$.  The result for $P_{\rm NLO}$ is
$T^4$ times a pure function of $\g2eff(\muDR)$,
\begin{equation}
P_{\rm NLO} = 2 \da \frac{\pi^2}{90} T^4 \; {\cal P}(\g2eff(\muDR)) \, .
\end{equation}
We have pulled out the free theory value; $2 \da=2 (\nc^2-1)$ counts the
number of degrees of freedom in the gauge field, and $\pi^2 T^4/90$ is
the usual result for a massless, bosonic field.

One may also compute the entropy density, which is the temperature
derivative of the pressure;
\begin{equation}
S_{\rm NLO} = \frac{dP_{\rm NLO}}{dT} = 2\da \frac{2 \pi^2}{45} T^3
	\left( {\cal P} + \frac{(\g2eff)^2}{24 \pi^2} \:
	\frac{d{\cal P}}{d \, \g2eff} \right) \, .
\end{equation}
This can be done from tabulated results for ${\cal P}$ by spline
interpolation, for instance.

\section{Results and Discussion}
\label{sec:results}

\begin{figure}
\centerline{\epsfxsize=3.2in\epsfbox{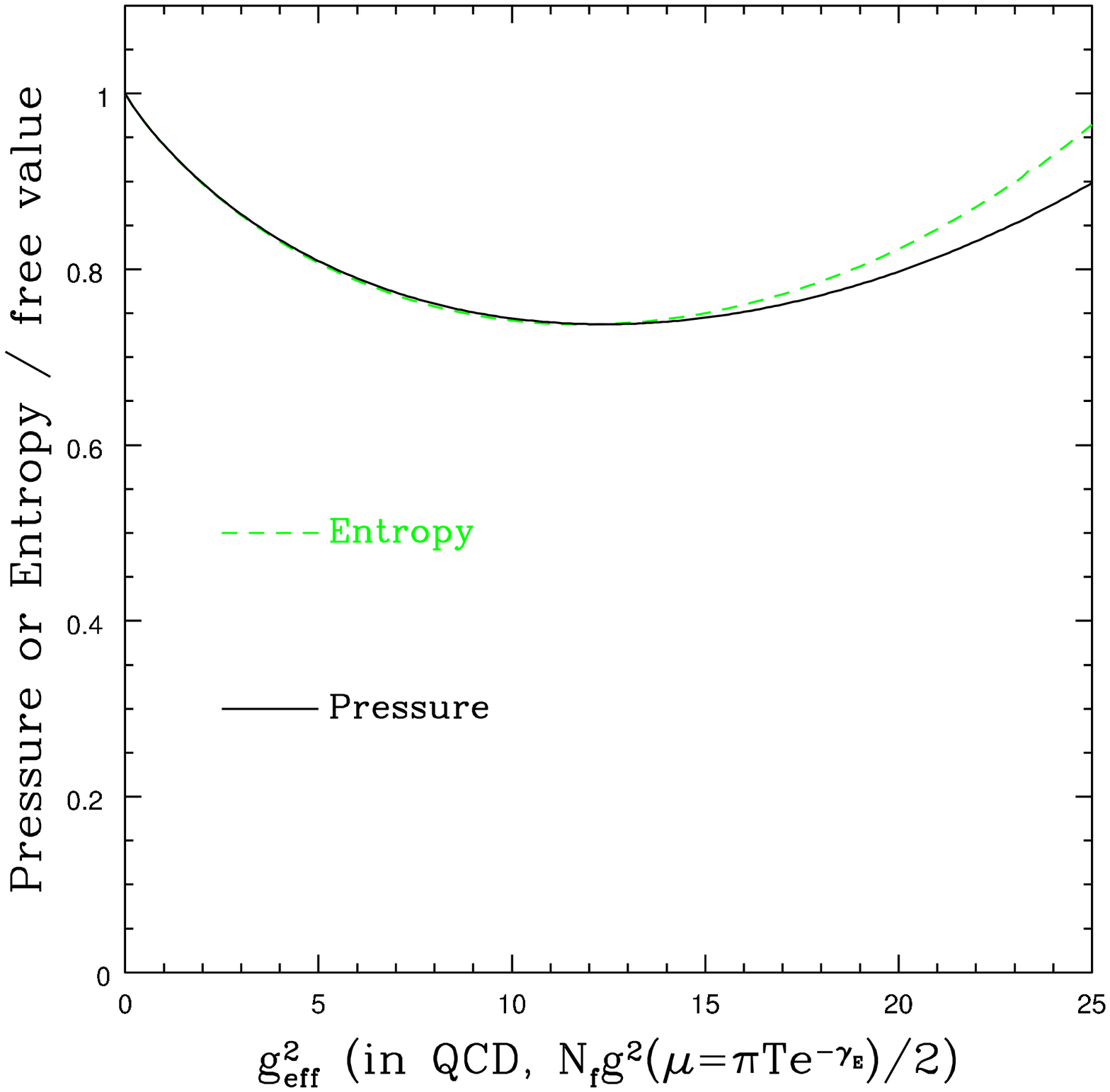} \hspace{0.2in}
\epsfxsize=3.2in\epsfbox{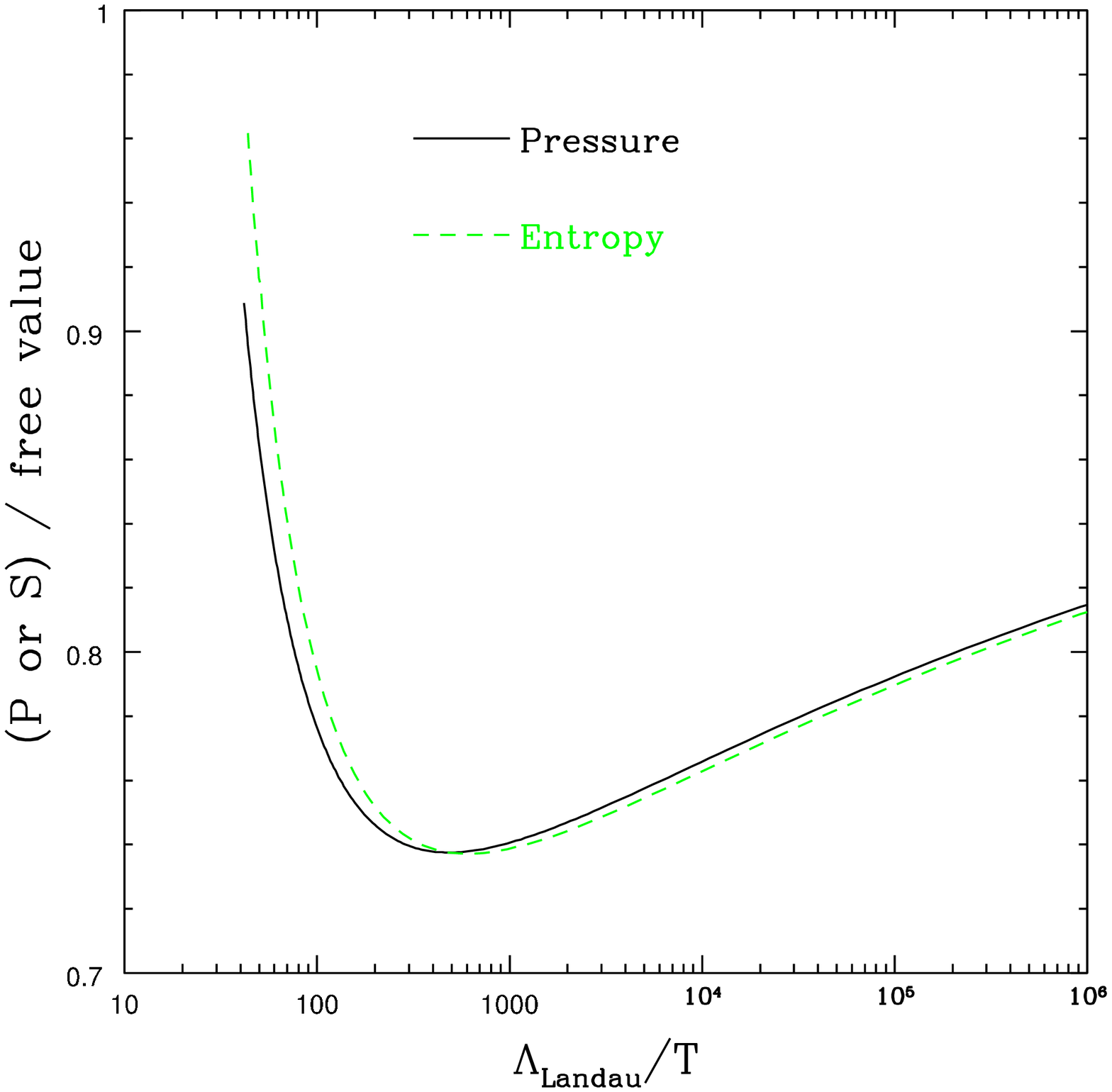}}
\caption{\label{fig:vs} ${\cal P} = P_{\rm NLO}/P_{\rm NLO, free}$ 
and $S_{\rm NLO} / S_{\rm NLO,free}$ plotted
against $\g2eff$, left, and $\LL/T$, right.
The right plot shows that the features indicating
the breakdown of perturbation theory--the pressure bottoming out and
then rising--occur long before the temperature reaches $\LL$.}
\end{figure}

\begin{figure}
\centerline{\epsfxsize=3.2in\epsfbox{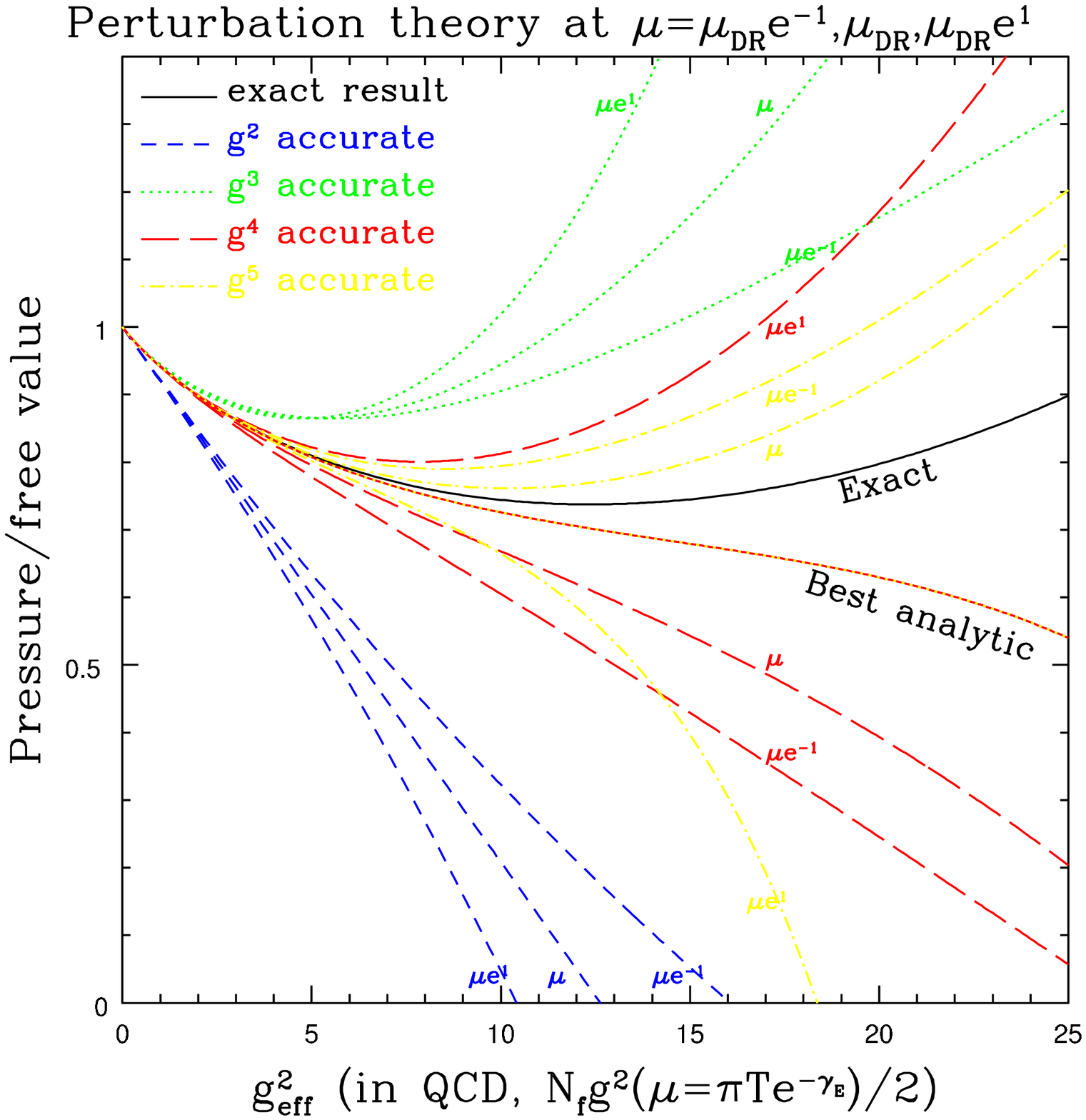} \hspace{0.2in}
\epsfxsize=3.2in\epsfbox{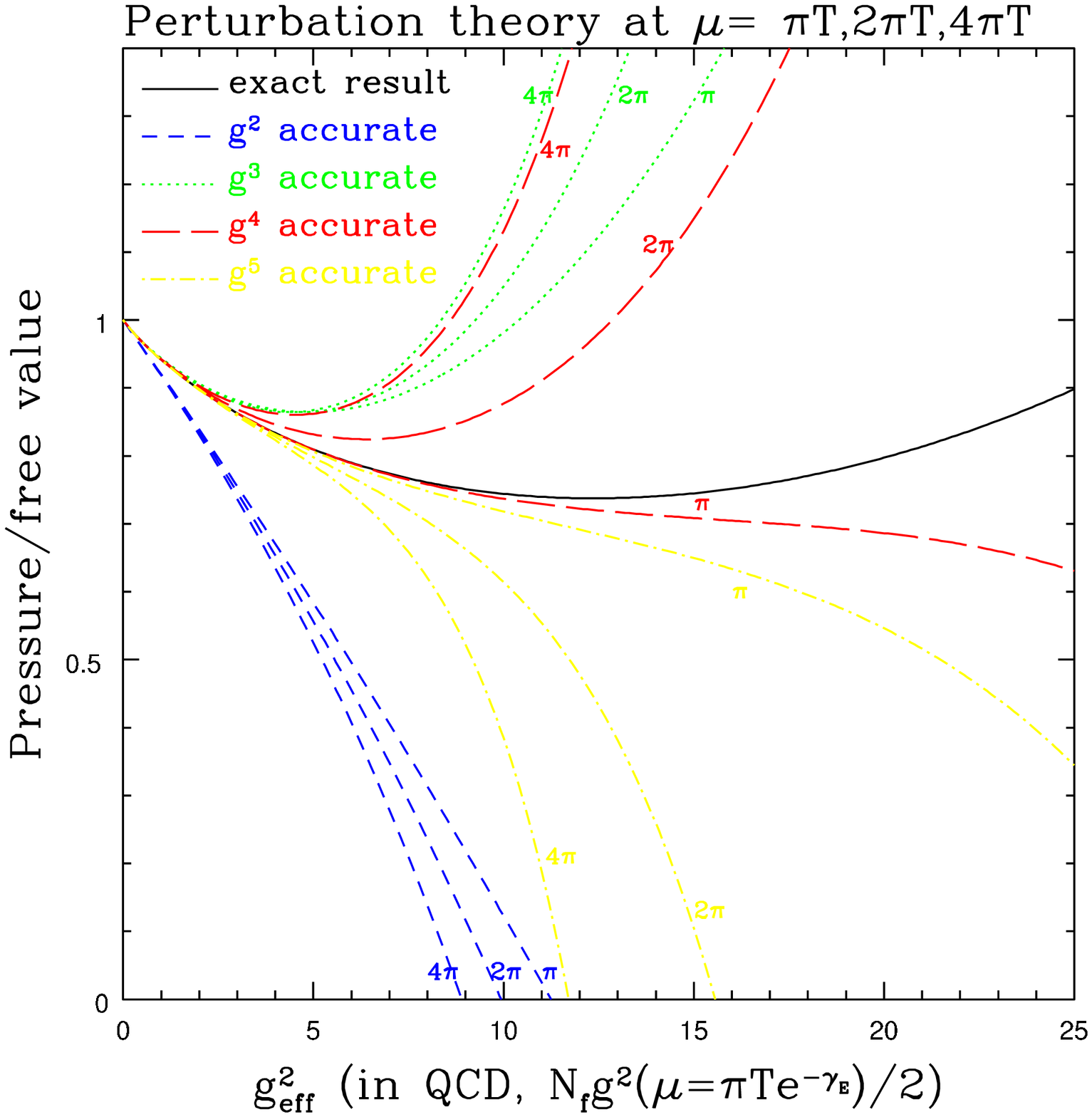}}
\caption{\label{fig:pert_thy} Exact result for $P_{\rm NLO}$, scaled by
free value, presented in comparison to the results of perturbation
theory.  Each figure shows three choices for the renormalization point
in perturbation theory; at left, they are within a power of $e$ of
$\muDR$; at right, they are within a factor of 2 of $2\pi T$.  The left
plot also shows the ``best'' perturbative result, as explained in the
text.}
\end{figure}

We present our results for the pressure and entropy, plotted against
$\g2eff$, in Fig.~\ref{fig:vs}.  The figure also shows pressure plotted
against $\LL/T$.  As the coupling is increased, ${\cal P}$ at first
falls, as expected (for instance in a quasiparticle picture); then it
flattens out and rises.  The figure shows that the minimum 
of the pressure occurs when
$\LL \simeq 480 T$.  This is a scale where the Landau singularity is
still far off in the ultraviolet; this minimum is a robust prediction of
the theory, regardless of the UV completion.

Note that in an earlier (and published!) version of this paper, the
numerical results were somewhat different (in particular, higher).  This
was a result of a coding error in evaluating \Eq{eq:P_minkowski}; to
take the imaginary part of the log, we used the arctangent of the
imaginary over the real piece.  For the longitudinal contribution, we
forgot to check whether the argument was in the principal range of the
arctangent function.  Andreas Ipp and Anton Rebhan pointed out a
discrepancy between the earlier result and their independent numerical 
evaluation of the expressions presented here \cite{RebhanIpp}, and I
thank them for correcting me.

These results are difficult to reconcile with
the ideal quasiparticle picture of
the plasma presented by Peshier, K\"{a}mpfer, Pavlenko, and Soff
\cite{Peshier_old}, in which the pressure and entropy are described by
treating the plasma as a gas of free but massive quasiparticles.
In this model it is unexpected for $S_{\rm NLO}/T^3$ to
rise as a function of $\g2eff$, because this requires that the
quasiparticle masses get smaller with increasing coupling.
Therefore
we are doubtful that an ideal quasiparticle picture can be a good
description of the physics of large $\nf$ QCD.  Since, as we have
argued, large $\nf$ is a subset of true QCD, we also don't expect it to
be a good description of QCD.

We also present the predictions of perturbation theory for the pressure,
plotted against the actual value.  The perturbative prediction for
$P_{\rm LO}$ agrees with \Eq{eq:LO}.  The NLO prediction, extracted
from the results of Zhai and Kastening \cite{ZhaiKastening}, is
\begin{eqnarray}
P_{\rm NLO} & = & 2 \da \: \frac{\pi^2 T^4}{90} \left\{
	1 - \frac{25}{2} \: \frac{\g2eff}{16 \pi^2}
	+ \frac{80}{\sqrt{3}} \: 
	\left( \frac{\g2eff}{16\pi^2} \right)^{3/2}
	+ \left[ \frac{100}{3} \ln \frac{\mu}{\muDR} - 68.907
	\right]	\left( \frac{\g2eff}{16\pi^2} \right)^{2}
	\right. \nonumber \\ && \hspace{0.7in} \left.
	+ \left[ \frac{-320}{\sqrt{3}} \ln \frac{\mu}{\muDR}
	+ \frac{160}{\sqrt{3}} \right]
	\left( \frac{\g2eff}{16\pi^2} \right)^{5/2}
	+ O\Big( (\g2eff)^3 \Big)
	\right\} \, .
\end{eqnarray}
Note that, unlike in QCD, there is in principle no obstacle to
computing the perturbative expansion to any order in $\g2eff$.  Also
there are no appearances of $\ln(\g2eff)$, because the 3-D dimensionally
reduced effective theory is free at order $\nf^0$.
We have evaluated each partial sum from this series, at a number of
renormalization points (running the coupling from $\muDR$ using the
exact relation, \Eq{eq:scale_dep}), in order to compare the perturbative
expansion (and its renormalization point sensitivity) to the exact
result.  The comparison is shown in Fig.~\ref{fig:pert_thy}.  The figure
also presents a ``best'' analytic result.  This is what one gets by
performing the dimensional reduction step to the highest order known
(two loop, in this case) and then solving the dimensionally reduced theory
exactly, as advocated recently by Laine \cite{Laine}.  The effective 3
dimensional theory can be solved exactly because it is a free theory.  The
procedure turns out to be equivalent to evaluating the perturbative
series above at the renormalization point where the $(\g2eff)^{5/2}$ term
vanishes.  As for
QCD, the perturbative expansion is valid at weak coupling but becomes
ill behaved long before the temperature reaches the Landau pole (or, in
QCD, the phase transition temperature).  The ``best'' analytic result
works well to surprisingly large coupling but eventually breaks down.

It is beyond the scope of the current paper to analyze large $\nf$ QCD
within each of the proposed resummation schemes, to compare to the exact
results presented here.  We leave this task to the
authors of those schemes.  However, to facilitate the
comparison, we present our numerical results for the pressure at a
number of values of $\g2eff$, in Table \ref{tab:1}.

\begin{table}
\begin{tabular}{|c|c|c|}
\hline
$\qquad \g2eff \qquad$ & ${\cal P}=P_{\rm NLO}/P_{\rm NLO}(\mbox{free})$ 
  & $\qquad \LL/T \qquad$ \\
\hline
0.0  & 1.00000   & $\infty$              \\ \hline
1.0  & 0.94163   & $2.1\times 10^{26}$   \\ \hline
2.0  & 0.89799   & $2.9 \times 10^{13}$  \\ \hline
3.0  & 0.86275   & $1.5 \times 10^{9} $  \\ \hline
4.0  & 0.83364    & $1.09 \times 10^{7} $ \\ \hline
5.0  & 0.8096    & $565,000$ \\ \hline
6.0  & 0.7897    & $78,000$ \\ \hline
7.0  & 0.7737    & $19,000$ \\ \hline
8.0  & 0.7610    & 6650 \\ \hline
9.0  & 0.7512    & 2920 \\ \hline
10.0 & 0.7442    & 1510 \\ \hline
11.0 & 0.7398    & 884 \\ \hline
12.0 & 0.7378    & 564 \\ \hline
13.0 & 0.7381    & 386 \\ \hline
14.0 & 0.7405    & 279 \\ \hline
15.0 & 0.7450    & 210 \\ \hline
16.0 & 0.7515    & 164 \\ \hline
17.0 & 0.7602    & 132 \\ \hline
18.0 & 0.7707    & 109 \\ \hline
19.0 & 0.7832    & 92 \\ \hline
20.0 & 0.797     & 78 \\ \hline
21.0 & 0.813     & 68 \\ \hline
22.0 & 0.832     & 60 \\ \hline
23.0 & 0.852     & 53 \\ \hline
24.0 & 0.874     & 48 \\ \hline
25.0 & 0.898     & 43 \\ \hline
\end{tabular}
\caption{ \label{tab:1} NLO pressure, normalized to the free value, as a
function of $\g2eff$.  The numerical error is
less than 2 in the last place shown, and is dominated by numerical
issues; Landau pole ambiguity is at most 1 in the last digit in all
entries.}
\end{table}

What are the implications of our results for full QCD?  Obviously it
does not make sense to use large $\nf$ QCD directly to model the
behavior of real world QCD; after all, in the physically interesting
range of temperatures, real world QCD has 3 flavors, which is definitely
not $\gg \nc = 3$.  This was anyway not the point of studying large
$\nf$ QCD.  Our object was to present a testing ground for other
techniques.  We expect these techniques to be accurate at small $\g2eff$
but, possibly, to break down at some larger value.  What does that imply
for their performance in real QCD?  As we have tried to emphasize,
success at large $\nf$ probably does not ensure success in full QCD; but
failure at large $\nf$ almost certainly means the technique is not
valid.  But it would also be nice to have a rough way of equating a
value of $\g2eff$ to a comparable value of the QCD coupling $\alphas$,
so if a technique works at weak coupling but later breaks down, we can
estimate its range of validity in full QCD.

Probably the most reasonable way to equate $\g2eff$ with $\alphas$
is to compare values of the Debye mass.  At least, this
makes sense in the picture where screening is the dominant physical
process changing the pressure.  The Debye mass for each theory, at
leading order, is%
\footnote
    {%
    had we chosen $\mubar = \muDR e^{1/2}$, the large $\nf$ expression
    would be correct through order $(\g2eff)^2$.}
\begin{equation}
\mD^2(\nf \gg 1) \simeq \frac{\g2eff}{3} T^2 \, , \qquad
\mD^2(\mbox{QCD}) \simeq \frac{2\nc + \nf}{6} g_{\rm s}^2 T^2 
	= \frac{2\pi(2\nc + \nf)}{3} \alphas T^2 \, .
\end{equation}
Equating these, for 3 flavor QCD, gives 
$\g2eff = 18 \pi \alpha_{\rm s}$.  So 
$\alphas = 0.11$ (relevant for temperatures of order 100GeV)
is comparable to $\g2eff = 6$, whereas $\alphas = .3$ (relevant closer
to the QCD phase transition temperature) is comparable to
$\g2eff = 17$.

In conclusion, we have presented exact results at next-to-leading order
(first nontrivial order) for the pressure of large $\nf$ QCD.  These
results should be useful for testing the reliability of resummed
perturbative techniques, used for full QCD.

\section* {ACKNOWLEDGMENTS}

I would like to thank Larry Yaffe for presenting this problem to me, and
for useful discussions; and Tony Rebhan, for correspondence on the
application of the 2PI method to this problem and for reading, and
catching an error in, the draft. 
I also thank Tony Rebhan and Andreas Ipp, for catching a numerical
error, which contaminated the original results presented in the first
version of this work. 
This work was supported, in part, by the U.S. Department
of Energy under Grant No.~DE-FG03-96ER40956 and by a McGill University
startup grant.

\appendix
\section{Self-energies}

The self-energies $\PiT$, $\PiL$ have been evaluated by Weldon
\cite{Weldon}; here we present his results, rotated into Euclidean
space, and present a few terms of the expansion about small $T^2/Q^2$.  

Following Weldon, we define two scalar functions, with normalization
chosen to agree with his usage,
\begin{eqnarray}
\label{eq:G}
2 \g2eff \; G(Q) & \equiv & \Pi^\mu_\mu(Q) - [\mbox{vac}] 
	= 2 \PiT(Q) + \PiL(Q) \, , \\
2 \g2eff \; H(Q) & \equiv & \Pi^{00}(Q) -[\mbox{vac}] = \frac{q^2}{Q^2} 
	\PiL(Q) \, ,
\end{eqnarray}
so that
\begin{eqnarray}
\PiL(Q) & = & 2 \g2eff \: \frac{Q^2}{q^2} H(Q) \, , \\
\PiT(Q) & = & \g2eff \left( -\frac{Q^2}{q^2} H(Q) + G(Q) \right) \, .
\end{eqnarray}
Weldon's results for these scalar quantities, rotated into Euclidean
space and for general $Q$, are
\begin{eqnarray}
G(Q) & = & \frac{1}{2\pi^2}\int_0^\infty \frac{dk}{e^{k/T}{+}1}
	\left[ 4k + \frac{Q^2}{2q} \ln 
	\frac{(q^0)^2 {+} (2k{-}q)^2}{(q^0)^2 {+} (2k{+}q)^2} \right] \, , \\
H(Q) & = & \frac{1}{2\pi^2} \int_0^\infty \frac{dk}{e^{k/T}{+}1}
	\left[ 2k + \frac{Q^2{-}4 k^2}{4q} \ln 
	\frac{(q^0)^2 {+} (2k{-}q)^2}{(q^0)^2 {+} (2k{+}q)^2} 
	\right. \nonumber \\ && \hspace{1.35in} \left.
	+ \frac{q^0 k}{q} 
	\: i \, \ln \frac{1 + 4k^2/(q^0{-}iq)^2}{1 + 4k^2/(q^0{+}iq)^2}
%
%\left( 2i \ln \frac{q^0{+}iq}{q^0 {-} iq}
%	+ i \ln \frac{4k^2 {+} (q^0{-}iq)^2}{4k^2 {+} (q^0{+}iq)^2}\right)
	\right] \, .
\end{eqnarray}
Both expressions are pure real, despite the appearance of $i$ in the
expression for $H(Q)$.  To obtain the retarded, Minkowski value,
replace $q^0 \rightarrow i\omega {+}0$.

It is possible to expand each quantity in $T^2 \ll Q^2$ by series
expanding the bracketed quantity in small $k$ and then performing the
$k$ integration.  This is expected to give an asymptotic series, since
the $k$ integration goes up to $\infty$ but with exponentially decaying
weight.  The next to leading order result, in Euclidean space, is
\begin{eqnarray}
\label{eq:G_expand}
G(Q\gg T) & \rightarrow
	& \frac{7 \pi^2 (3 (q^0)^2 - q^2)}
	{45 (Q^2)^2} T^4
	- \frac{248 \pi^4(5 (q^0)^4 - 10(q^0)^2 q^2 + q^4)}
	{315 (Q^2)^4} T^6 + O(T^8)
	\, , \\
H(Q\gg T) & \rightarrow
	& \frac{7 \pi^2 q^2}{45 (Q^2)^2} T^4
	- \frac{248 \pi^4 q^2 (5(q^0)^2 - q^2)}{945 (Q^2)^4} T^6
	+O(T^8)
	\, .
\end{eqnarray}

\end{document}